\documentclass{aastex}
\usepackage{emulateapj5}
\usepackage{apjfonts}
\usepackage{epsf}

\def\ah{{\cal A_{\rm H}}}
\def\ahe{{\cal A_{\rm He}}}
\def\afe{{\cal A_{\rm Fe}}}
\def\cc{{\cal C}}

\def\ergs{{\rm erg~s^{-1}}}

\def\pe{\tilde{\epsilon}}

\def\pp{\prime\prime}

\begin{document}

\title{The Afterglows of Ultraluminous Quasars}

\author{Jian-Min Wang\altaffilmark{1}, 
Ye-Fei Yuan\altaffilmark{2} and 
Luis C. Ho\altaffilmark{3}}

\altaffiltext{1}{Laboratory for High Energy Astrophysics, Institute of High Energy Physics,
Chinese Academy of Sciences, Beijing 100039, China; wangjm@mail.ihep.ac.cn}

\altaffiltext{2}{
Center for Astrophysics, University of Science and Technology of China, Hefei 230026, China;
yfyuan@ustc.edu.cn}

\altaffiltext{3}{
The Observatories of the Carnegie Institution of Washington, 813 Santa Barbara Street, Pasadena,
CA 91101, USA; lho@ociw.edu}

\slugcomment{Accepted by The Astrophysical Journal Letters}
\shorttitle{Afterglows of Ultraluminous Quasars}
\shortauthors{WANG, YUAN \& HO}

\begin{abstract} 
Quasars represent a brief phase in the life-cycle of most massive galaxies, but
the evolutionary connection between central black holes and their host 
galaxies remains unclear.  While quasars are active and shining brighter than 
the Compton-limit luminosity, their radiation heats the surrounding medium to 
the Compton temperature, forming Compton spheres extending to the Str\"omgren 
radius of Fe$^{26+}$/He$^{2+}$.  After the quasars shut off, their 
``afterglow'' can be detected through three signatures: (1) an extended X-ray 
envelope, with a characteristic temperature of $\sim 3\times 10^7$ K; (2) 
Ly$\alpha$ and Ly$\beta$ lines and the $K-$edge of Fe$^{26+}$; and 
(3) nebulosity from hydrogen and helium recombination emission lines.  
We discuss the possibility of detecting these signatures using 
{\em Chandra}, the planned {\em XEUS} mission, and ground-based optical telescopes.
The luminosity and size of quasar afterglows can be used to constrain the 
lifetime of quasars.
\end{abstract}

\section{INTRODUCTION}
The neutral hydrogen bounded by dark matter halos at high redshift has 
detectable extended Ly$\alpha$ ``fuzz'' powered by photoionization from 
central quasars (Haiman \& Rees 2001). The detection of this signal may impose 
strong constraints on quasar formation during galaxy assembly.  The lifetime of 
quasars remains uncertain, but they are generally thought to be short-lived, 
with estimated ages of $\sim 10^7$ to $10^9$ years (Jakobsen et al. 2003; Martini 
\& Schneider 2003).  
Detecting the infalling gas, heated by the quasar, as Ly$\alpha$ fuzz can help 
to constrain the quasar's lifetime.  If undetected, it may imply that quasar 
activity only appears when the gas has settled into a thin disk or has already 
been consumed mostly by star formation. The preliminary evidence for extended 
Ly$\alpha$ emission in radio-quiet quasars already suggests that quasar 
activity influences galaxy assembly  (Bergeron et al. 1999; Fardal et al. 2001;
Bunker et al. 2003). The spectrum of the quasar Q1205$-$30 unambiguously shows 
an extended hydrogen Ly$\alpha$ glow formed by gas falling into the ionizing 
cone of the quasar (Weidinger et al. 2004).

What is the physical appearance of a quasar turning {\it off}\ within an 
assembling galaxy?  This Letter investigates the afterglow properties of 
ultraluminous, radio-quiet quasars.

\section{COMPTON SPHERE AND COMPTON LIMIT}
While the quasar is active, it provides a source of energy to heat infalling 
material (Binney \& Tabor 1995; Ciotti \& Ostriker 2001; Brighenti \& Mathews 2002). 
Hot gas in elliptical galaxies has been extensively studied 
for three decades since {\em Einstein}, but the physics of its origin and evolution 
still are not completely understood (see the review by Mathews \& Brighenti 2003).
It has been realized that gas infalling into the  dark matter halo 
of a nascent galaxy (Fall \& Rees 1985) undergoes thermal instability and forms a 
two-phase medium that would exhibit very faint emission features (Haiman, Spaans \&
Quataert 2000). In this stage, pressure balance holds between the cold 
($T_{\rm c} \approx 10^4$ K) and hot ($T_{\rm h} \approx 10^6$ K) medium. The clumping 
factor of the two-phase medium is $\cc=\langle n_{\rm b}^2\rangle/\langle 
n_{\rm b}\rangle^2=\left(T_{\rm h}/T_{\rm c}\right)^2\approx 10^4$, where
$n_{\rm b}$ is the baryon number density. 
The irradiation by the quasar light greatly changes the two-phase structure of 
the infalling gas.  We assume the abundance of hydrogen, helium and iron to be 
$\ah=0.7$, $\ahe=0.3$ and $\afe=5.0\times 10^{-5}$, respectively. The thermal 
state of the photoionized gas is determined by the ionization parameter, 
defined by 
\begin{equation}
\Xi=\frac{F}{cnkT}, 
\end{equation}
where $n$ and $T$ are the number density and 
temperature of the gas and $F$ is the flux of the ionizing source (Krolik, McKee \& Tarter
1981). Compton heating and cooling balance at a temperature of $T_{\rm C}=\pe/4k$, 
where $\pe$ is the mean energy of the ionizing source and $k$ is Boltzmann's 
constant. For a typical quasar spectrum (Elvis et al. 1994), the mean photon energy 
is $\pe\approx 10$ keV. The gas will be heated to $T_{\rm C}\approx 
3.0\times 10^7$ K (see the detailed calculation of Sazonov et al. 2004), which is one 
order of magnitude hotter than the typical 
virial temperature of 0.15 keV for the hot gas (Fall \& Rees 1985). When the ionization 
parameter of the ionized gas reaches 
\begin{equation}
\Xi\ge \Xi_{\rm C}=6.1\pe_1^{-3/2}, 
\end{equation}
where $\pe_1=\pe/10\,{\rm keV}$, the gas will be fully heated to the Compton 
temperature (Krolik, McKee \& Tarter 1981). The timescale of Compton heating is given 
by $t_{\rm C}^{\rm heat}=3m_ec^2kT_{\rm C}/8\sigma_{\rm T}\pe F$, where $m_e$ is 
the electron mass and $\sigma_{\rm T}$ is the Thomson scattering cross section. This
timescale is derived by dividing the number of scatterings needed to reach the Compton 
temperature by the scattering rate.
We assume a quasar lifetime of $t_{\rm Q}=\tau_{\rm Q}t_{\rm Salp}$, with 
$t_{\rm Salp}=4.6\times 10^8$ yr being the Salpeter 
timescale (Salpeter 1964).  Setting $t_{\rm C}^{\rm heat}=t_{\rm Q}$, with 
$L_{46}=L/10^{46}\,\ergs$, we have the 
Compton radius
\begin{equation}
R_{\rm C}=3.2~ L_{46}^{1/2}\tau_{\rm Q}^{1/2}~{\rm kpc},
\end{equation}
which, along with the condition $\Xi=\Xi_{\rm C}$, gives the Compton density 
\begin{equation}
n_{\rm C}=1.1\times 10^{-2}~ \tau_{\rm Q}^{-1}\pe_1^{1/2}~~~{\rm cm^{-3}}.
\end{equation}
%
\figurenum{1}
\centerline{\includegraphics[angle=-90,width=7.5cm]{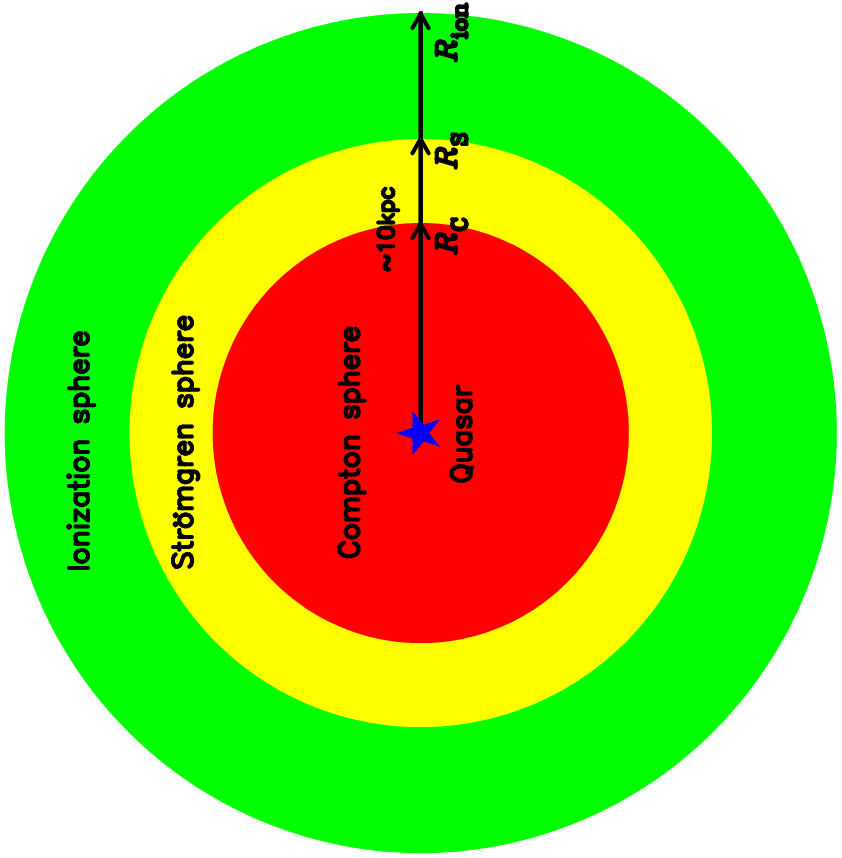}}
\vglue 0.2cm
\figcaption{\footnotesize  
Illustration of the afterglow from an ultraluminous quasar. The Compton
radius, $R_{\rm C}$, extends to the Str\"omgren radius, $R_{\rm S}$,  during
the quasar's lifetime when its luminosity exceeds the Compton limit luminosity.
The afterglow arises from the Compton sphere after the quasar is extinct.}
\label{fig1}
\vglue 0.5cm
%
The Compton radius depicts the size of the region in which the gas is heated 
by the quasar to the Compton temperature during the timescale $t_{\rm Q}$.
The Compton density is the maximum density allowed within $R_{\rm C}$ by 
Compton heating during the lifetime of the quasar.  It is interesting to note 
that $n_{\rm C}$ is independent of the quasar luminosity.  Beyond $R_{\rm C}$, 
the gas has an ionization parameter of $\Xi< \Xi_{\rm C}$ and hence is in a 
two-phase state. For a medium with a density lower than $n_{\rm C}$, the size 
of the region with $\Xi>\Xi_{\rm C}$ extends beyond $R_{\rm C}$. However, in 
such a case Compton heating is sufficiently inefficient that the 
gas is in a time-dependent two-phase state.  We do not consider this 
regime for ultraluminous quasars.

The radius of the Str\"omgren sphere of Fe$^{26+}$ ions is 
given by
\begin{equation}
R_{\rm S}^{\rm Fe XXVI}=\left(\frac{3\dot{N}_{\rm ion}}{4\pi \alpha_{\rm Fe}
               {\cal A}_{\rm Fe}\cc\langle n_b^2\rangle}\right)^{1/3}
     =5.7~L_{46}^{1/3}\cc_4^{-1/3}\langle n_b^2\rangle_{-4}^{-1/3}~{\rm kpc},
\end{equation}
where $\cc_4=\cc/10^4$, $\alpha_{\rm Fe}=9.06\times 10^{-12}~{\rm cm^3~s^{-1}}$
is the Fe$^{26+}$ recombination coefficient in Seaton (1959) evaluated at 
$T_e=3\times 10^7$ K, and 
$\langle n_b^2\rangle_{-4}=\langle n_b^2\rangle/10^{-4}~{\rm cm^{-6}}$ is the 
volume-averaged mean square density of the cold gas. The rate of ionizing 
photons, $\dot{N}_{\rm ion}$, is estimated from the typical spectrum of 
quasars (Elvis et al. 1994).  The radius of the He$^{2+}$ Str\"omgren sphere
$R_{\rm S}^{\rm He III}=10.3~L_{46}^{1/3}\cc_4^{-1/3}\langle n_b^2\rangle_{-4}^{-1/3}~{\rm kpc}$, 
and $R_{\rm S}^{\rm He III}/R_{\rm S}^{\rm H II}\approx 0.3$;
we assume ionization energies of $E_{\rm He III}=54.4$ eV and 
$E_{\rm H II}=13.6$ eV and an ionizing spectrum $f_\nu \propto \nu^{-1.3}$. 
The radius of the ionized region is given by (Cen \& Haiman 2000)
$R_{\rm ion}^{\rm He III}=\left(3\dot{N}_{\rm ion}t_Q/4\pi\ahe\langle n_b^2\rangle^{1/2}\right)^{1/3}
           =119~\tau_{\rm Q}^{1/3}L_{46}^{1/3}\langle n_b^2\rangle_{-4}^{1/6}~~{\rm kpc}$, and
$R_{\rm ion}^{\rm He III}/R_{\rm ion}^{\rm H II}=1.2$.
As the ionizing luminosity increases, the Compton radius extends to the radius of the 
helium Str\"omgren sphere. The entire Str\"omgren sphere will be fully heated 
to the Compton temperature when $R_{\rm S}^{\rm Fe XXVI} = R_{\rm C}$, at 
which point the iron Compton-limit luminosity
\begin{equation}
L_{\rm C}^{\rm Fe}=3.0\times 10^{47}~\tau_{\rm Q}^{-3}\cc_4^{-2}\langle n_b^2\rangle_{-4}^{-2}
             ~~\ergs,
\end{equation}
and the iron Compton-limit radius 
\begin{equation}
R_{\rm C}^{\rm Fe}=17.7~\tau_{\rm Q}^{-1}\cc_4^{-1}\langle n_b^2\rangle_{-4}^{-1}~~{\rm kpc}.
\end{equation}
%
\figurenum{2}
\centerline{\includegraphics[angle=-90,width=7.5cm]{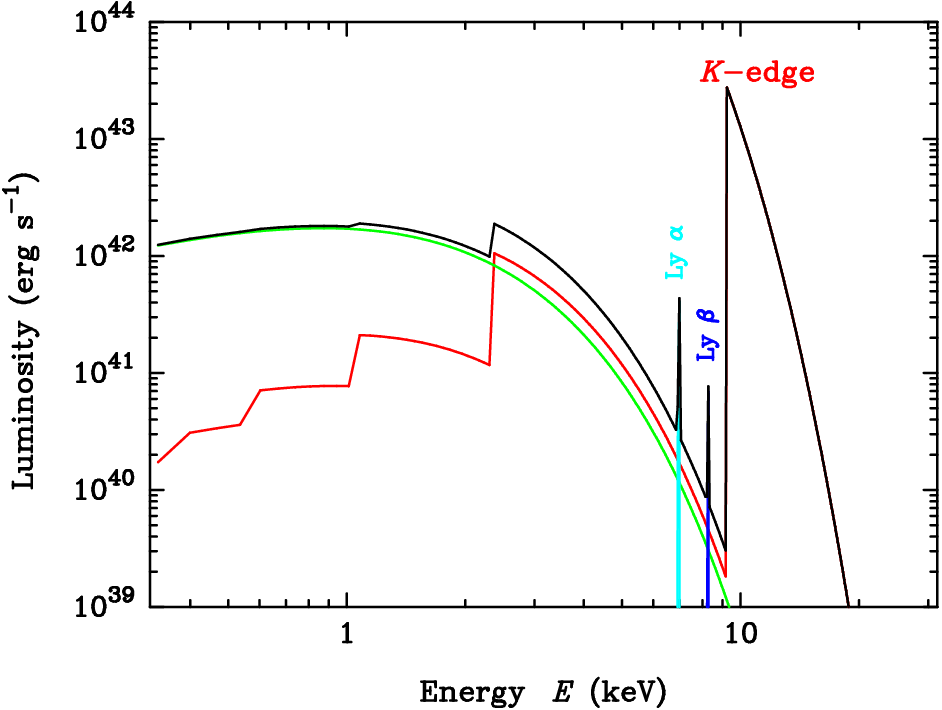}}
\vglue 0.2cm
\figcaption{\footnotesize  
The X-ray spectrum of the Fe$^{26+}$ Compton sphere. The green line represents
free-free emission, the red line free-bound (radiative recombination) emission,
and the black line the total emission from the Compton sphere. Iron Ly$\alpha ~\lambda1.78$ 
and Ly$\beta ~\lambda1.51$ lines, with emissivities (Raymond \& Smith 1977) for
$\sim 3\times 10^7$ K and solar iron abundances, are labeled
($\Delta E/E=0.02$ for {\em XEUS}).
The Ly$\alpha$ and Ly$\beta$ lines will have equivalent widths of
12.0 keV and 6.6 keV for solar iron abundances, and should be detectable with
{\em XEUS}.}
\label{fig2}
\vglue 0.2cm
When $L<L_{\rm C}^{\rm Fe}$, the Compton sphere is smaller than the 
Str\"omgren sphere.  Since the medium in the region between the Str\"omgren 
radius and the Compton radius cannot fully reach the Compton temperature, the 
features of the afterglow arising from it could be complicated.  We are 
currently interested in the case of $L\ge L_{\rm C}^{\rm Fe}$, namely, the 
ultraluminous quasars. 

The energy stored within the Compton sphere, 
$\frac{4}{3}\pi R_{\rm C}^3n_ekT\approx 4\times 10^{58}n_{-2}$ erg, where 
$T=3.0\times 10^7$ K, $n_{-2}=n_e/10^{-2}$ cm$^{-3}$, and $R_{\rm C}=20$ kpc, is 
only a small fraction ($0.02\%$) of the total energy released 
($2\times 10^{62}$ erg) during the activity of a quasar with a black hole mass 
of $10^9\, M_{\odot}$ and an accretion efficiency of 0.1.  The recombination 
timescale is $(\alpha_{\rm Fe}n_e)^{-1}\approx 0.3\times 10^7n_{-2}^{-1}$ yr.  For 
quasars surrounded by a very dilute medium, this timescale is too long for the 
ions to recombine within the Hubble time at the relevant epoch.  The 
Str\"omgren radius is then never reached, and the hot gas would emit no 
recombination lines. This situation might be relevant for some isolated, 
exceptionally gas-poor elliptical galaxies.  If the recombination timescale is 
longer than the lifetime of quasars, the discussion on the Str\"omgren radius 
is also inapplicable.

\section{OBSERVATIONAL SIGNATURES}
An afterglow will be seen from the Compton sphere after the quasar is extinct. 
Here we estimate the expected signature of the quasar afterglow from the hot 
gas inside the Compton sphere that is in hydrostatic equilibrium with the 
gravitational potential of the dark matter halo (Fig. 1). The isothermal density 
profile of the hot gas is given by (Makino et al. 1998) 
$\rho_{\rm g}=\rho_0 e^{-\beta}\left(1+R/R_{\rm s}\right)^{\beta R_{\rm s}/R}$, 
where $\rho_0=1.96\times 10^{-26}~{\rm g~cm^{-3}}$, $\beta=3.03M_{13}T_7^{-1}$,
the characteristic radius $R_{\rm s}=44.25M_{13}^{1/3}~{\rm kpc}$, 
$T_7=T/10^7~{\rm K}$, and $M_{13}=M_{\rm halo}/10^{13}~M_{\odot}$. We use a 
concentration parameter of $c_0=5$ and $\Delta_{\rm c}=18\pi^2$ for the dark 
matter halo. The gas mass fraction is only $\sim 0.005(\tau_{\rm es}/10^{-3})$ 
of that of the dark matter halo, where $\tau_{\rm es}$ is the Thomson 
scattering depth of the hot gas. We note that the density of the hot gas is 
lower than the Compton density.

The cooling timescale of the fully ionized gas inside the Compton sphere is 
governed mainly by free-free radiation.  
%
\figurenum{3}
\centerline{\includegraphics[angle=-90,width=7.5cm]{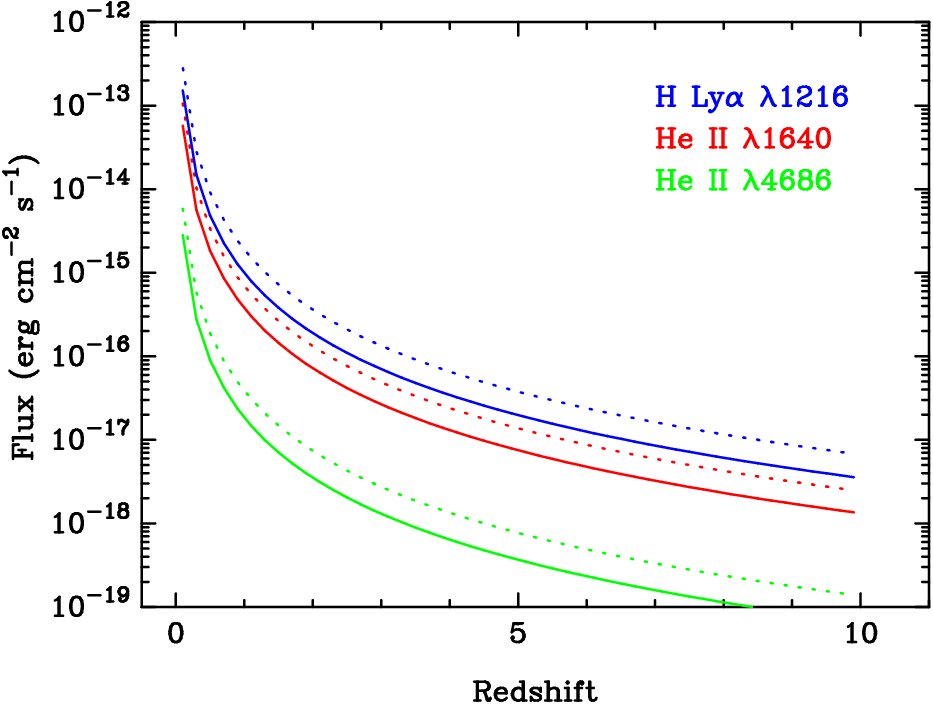}}
\vglue 0.2cm
\figcaption{\footnotesize  
Expected fluxes of hydrogen (Ly$\alpha$ $\lambda$1216) and helium
(He~{\sc ii} $\lambda\lambda$1640, 4686) recombination lines. The solid and
dotted lines show the predictions for an electron temperature of
$10^4$ K and $2\times 10^4$ K, respectively.
We assume a $\Lambda$CDM cosmology with $\Omega_{\rm m}=0.3$,
$\Omega_{\Lambda}=0.7$, and $H_0=75$~km~s$^{-1}$~Mpc$^{-1}$.}
\label{fig3}
\vglue 0.2cm
The cooling timescale at the center 
of the halo is $t_{\rm ff} \approx 1.0\times 10^9T_7^{1/2}n_{-2}^{-1}$ yr, 
and it will be considerably longer in the outskirts. This leads to the 
appearance of a long-lived, extended X-ray--emitting envelope, characterized by 
a luminosity $L_{\rm ff}\approx 1.3\times 10^{42}~\ergs$ for $T=10^7$ K,  
$R_{\rm C}=20$ kpc and the above density profile,  peaking at an energy 
of a few keV.  Since the degree of ionization is very high during the active 
phase of the quasar, we assume that iron is fully ionized.  Figure 2 shows the
predicted X-ray signature from such a model, where we have computed the 
integrated emission from the Compton sphere of a progenitor quasar shining at 
a fiducial Compton luminosity of $10^{42}~\ergs$, including contributions from 
the $K$ and $L$ edges of Fe$^{26+}$ free-bound emission from hydrogen-like 
ions (Seaton 1959) as well as the Ly$\alpha$ and Ly$\beta$ lines from Fe$^{26+}$.  
For redshifts $z\le 0.2$, the expected flux of 
$\ge 1.1\times 10^{-14}~{\rm erg~s^{-1}~cm^{-2}}$ will be detectable 
by {\it Chandra}/ACIS with an exposure time of 100 ks, and the angular radius 
of the Compton sphere, $\theta\approx 29.5^{\pp}$, should be resolvable.
However, the effective area of ACIS ($\sim 80~{\rm cm^2}$ above a few keV)
is too small to be able to detect the Ly$\alpha$ and Ly$\beta$ lines of 
Fe$^{26+}$. The future {\em XEUS} mission, with an expected angular resolution 
of $2^{\pp}$ at a flux level of 
$10^{-17}~{\rm erg~s^{-1}~cm^{-2}}$ in the energy range $0.05-30$ keV, will be
able to detect the Compton sphere and Ly$\alpha$ and Ly$\beta$ emission at 
redshifts lower than 1.  Higher-redshift afterglows are too faint to be 
detected by {\em Chandra}, whereas {\em XEUS} will have insufficient resolution.

Lastly, we consider the signal from the region between the Compton radius and 
the ionization radius (Fig. 1), which arises from hydrogen and helium 
photoionized by the quasar. After the quasar quenches, the photoionized gas
will radiate recombination lines of hydrogen and helium. We calculate the 
fluxes of hydrogen Ly$\alpha$ and He~{\sc ii} $\lambda\lambda$1640, 4686, 
using  the recombination coefficients in Seaton (1978), by 
integrating the shell between the Compton and ionization radius (Fig. 3).
The broadening width of the recombination lines due to thermal motions,
$\Delta \lambda/\lambda\approx \left(kT_{\rm C}/Zm_pc^2\right)^{1/2}\approx 
(0.2-3.0)\times 10^{-3}$, where $Z$ is the charge of the ions.  This width 
can be readily resolved by ground-based optical telescopes, the largest of 
which should also have the requisite sensitivity and spatial resolution to 
image the line-emitting region.
We note that the emission-line regions considered here are different from the 
extended Ly$\alpha$  nebulae discovered in some star-forming 
galaxies (Steidel et al. 2000). The latter are powered by OB stars, which are 
not energetic enough to ionize He.

If quasar afterglows can be detected, either the luminosity or the size of the 
Compton sphere (Equations 6 and 7) can be used to estimate $\tau_{\rm Q}$, 
the lifetime of the progenitor quasar.  
In practice, the luminosity would be much better constrained than the size, 
as the latter would be only marginally resolved even with {\em Chandra} and 
depends critically on the detailed surface brightness distribution.  While the 
hot gas normally associated with individual elliptical galaxies (O'Sullivan et al.
2003) rarely has temperatures exceeding 1~keV, galaxies situated in groups 
or denser environments would be surrounded by hotter gas, which at large 
distances would be difficult to distinguish from any component intrinsic to an 
afterglow.  To minimize this source of confusion, the local environment of any 
potential targets needs to be carefully scrutinized.  With typical X-ray 
temperatures of a few keV,
X-ray binaries present another possible source of contamination, but their
integrated X-ray emission (Fabbiano 1989), even in large galaxies, is 
significantly less than $10^{42}~\ergs$.

If quasars shine near the Eddington limit, the Compton-limit luminosity 
corresponds to a black hole mass $M_{\rm BH}\approx 2\times 10^9~M_{\odot}$. 
If such massive black holes are common among quasars (Netzer 2003; Vestergaard 
2004), their afterglows should not be difficult to detect. From the 
$M_{\rm BH}-\sigma$ relation (Tremaine et al. 2002), we expect the host 
galaxies to have stellar velocity dispersions of $\sigma \approx 
400~{\rm km~s^{-1}}$. Thus, future searches of quasar afterglows should target 
the most massive giant ellipticals. Such galaxies are rare, but not exceedingly
so.  From the local velocity dispersion function of early-type galaxies 
(Sheth et al. 2003), the space density of galaxies with $\sigma\approx 
400~{\rm km~s^{-1}}$ is $\sim 10^{-6}$ Mpc$^{-3}$.
 
For the more abundant but less powerful quasars, with luminosities below the 
Compton-limit luminosity, the Compton radius cannot match the Str\"omgren 
radius and it will be significantly smaller than the Compton-limit radius. The 
medium in the transition zone between the Compton and Str\"omgren radii is in a 
time-dependent two-phase state, since it is not able to reach an equilibrium 
state within the lifetime of the quasar. We speculate that these 
lower-luminosity sources would have optical and ultraviolet emission features 
quite similar to those of ultraluminous quasars, but that the X-ray spectrum 
should be dominated by the Ly$\alpha$ and Ly$\beta$ lines of lighter ions, 
such as Ne~{\sc x} or O~{\sc viii}.  Detailed calculations for this case 
are in progress.

\section{SUMMARY}
In conclusion, we show that the afterglows of ultraluminous quasars form 
relic Compton spheres that could be detected by {\em Chandra}, {\em XEUS}, and 
ground-based optical telescopes.  The luminosities and sizes of the 
afterglows can be used to estimate the lifetime of quasars, a key constraint on
quasar and galaxy evolution.  
The extended X-ray envelope recently detected around the 
isolated elliptical galaxy NGC 4555 (O'Sullivan \& Ponman 2004) could be an example 
of the afterglows discussed in the present paper.  It would be worthwhile to search 
in this object for the predicted extended emission from hydrogen and helium 
recombination lines.

\acknowledgements 
We thank the referee, Dr. Markos Geoganopoulos, for helpful comments.
This research is supported by a Grant for Distinguished Young
Scientists through NSFC-10325303, NSFC-10222030 and 973 Projects.  L.C.H. 
thanks the Carnegie Institution of Washington for support and J. S. Mulchaey 
for useful discussions.

\clearpage

\clearpage

\clearpage

\end{document}